\documentclass[12pt]{article} 
\usepackage{titlesec}

\titleformat*{\section}{\large\bfseries}
\titleformat*{\subsection}{\normalsize\bfseries}

\usepackage{epsf}
\usepackage[dvips]{graphicx}
\usepackage{amssymb}

\begin{document} 

\begin{center} 

{\large\bfseries 
Self-Similarity, Fractality and Entropy Principle \\
in Collisions of Hadrons and Nuclei \\ 
at Tevatron, RHIC and LHC \\}

\vskip 8mm
I. Zborovsk\'{y}$^{\star, \flat}$ and M. Tokarev$^{\star\star, \natural}$ 

\vskip 5mm
{\small ($\star$) {\it The Czech Academy of Science, Nuclear Physics Institute, 
\v{R}e\v{z}, Czech Republic }}
\\
($\star\star$) {\it Joint Institute for Nuclear Research,
Dubna, Russia }
\\
$\flat$ {\it E-mail: zborovsky@ujf.cas.cz }
\\
$\natural$ {\it E-mail: tokarev@jinr.ru } 

\end{center}     

\vskip 8mm
\abstract{
$z$-Scaling of inclusive spectra as a manifestation of self-similarity and 
fractality of hadron interactions is illustrated.
The scaling  for negative particle production in $Au+Au$ collisions
from BES-I program at RHIC is demonstrated. The scaling variable $z$ depends 
on the momentum fractions of the colliding objects carried by the 
interacting constituents, and on the momentum fractions of the fragmenting objects  
in the scattered and recoil directions carried by
the inclusive particle and its counterpart, respectively.  
Structures of the colliding  objects and fragmentation processes in final state
are expressed by fractal dimensions.  
Medium produced in the collisions is described by a specific heat. The scaling 
function $\psi(z)$ reveals energy,  angular,  multiplicity,  and flavor independence.
It has a power behavior at high $z$ (high $p_T$). Based on the entropy principle 
and $z$-scaling, energy loss as a function of the collision energy, 
centrality and transverse momentum of inclusive particle is estimated.
New conservation law including fractal dimensions is found. 
Quantization of fractal dimensions is discussed.
}

\vskip 7cm
\noindent
{\it
40th International Conference on High Energy Physics,\\
28 July - 6 August, 2020, Prague,  Czech Republic}

\hspace{-0.6cm}\rule{9.5cm}{0.4pt}

$\star$ {Speaker}

\newpage

\section{Introduction}

The production of particles with high transverse momenta from the collisions of
hadrons and nuclei at sufficiently high energies has relevance to constituent interactions
at small scales. In this regime, it is interesting to search for new physical
phenomena in elementary processes such as quark compositeness \cite{1n,2n},
extra dimensions \cite{3n}-\cite{7n}, black holes \cite{8n}-\cite{10n}, 
fractal space-time \cite{11n}-\cite{13n}, fundamental symmetries \cite{Lederman}, etc. 
Other aspects of high energy 
interactions are connected with small momenta of secondary particles and high
multiplicities. In this regime, collective phenomena of particle production take
place. Search for new physics in both regions is one of the main goals of investigations
at the Relativistic Heavy Ion Collider (RHIC) at BNL  
and the Large Hadron Collider (LHC) at CERN. 
Processes with high transverse momenta of 
produced particles are most suitable for a precise test of perturbative Quantum
Chromodynamics (QCD). 
The soft regime is preferred for verification of nonperturbative  
QCD and investigation of phase transitions in non-Abelian theories.
Basic principles which lie in the root of modern physical theories are the principles
of relativity (special, general, scale), gauge invariance, locality, spontaneous
symmetry breaking and others. The fundamental property of asymptotic freedom
and gauge invariance were used for the development of QCD, the theory of strong
interactions of quarks and gluons. 
Due to asymptotic freedom,
the perturbative QCD is controlled by higher-order corrections in strong coupling constant.
However, there is no universal method to be used in the nonperturbative sector of QCD. 
This is a big challenge because even at large momentum transfer, some nonperturbative
aspects of the theory are still needed to make a comparison with measured physical
observables. Moreover, the internal 
structure of hadrons is not fully understood
and remains largely mysterious especially at small scales. In such situation,
the principles of self-similarity and fractality can give additional constrains for
theories in particle physics. We consider that both mentioned principles reflect
general features of hadron interactions at high energies. They are important for
verification of symmetries already established, study of their possible violations, and 
search for new symmetries which govern physical theories.

\section{Scaling and universality as general concepts}

The idea of self-similarity of hadron interactions is a fruitful concept 
to study collective phenomena in hadron matter.
Important manifestation of such a concept 
is existence of scaling itself (see \cite{Stanley1}-\cite{Domb} and references therein). 
Scaling in general means self-similarity at different scales. 
The physical content meant behind it can be of different origin. 
Some of the scaling features constitute pillars of modern critical phenomena. 
Other category of scaling laws (self-similarity in point explosion, laminar and turbulent 
fluid flow, super-fluidity far from phase boundary and critical point, etc.) 
reflects features not related to phase transitions.

The notions "scaling" and "universality" have special importance in critical phenomena. 
%The concepts developed to understand the critical phenomena are "scaling" and "universality". 
The scaling means that the system near the critical point exhibiting self-similar properties 
is invariant under transformation of a scale. According to universality, quite different 
systems behave in a remarkably similar way near the respective critical point.  
The universality hypothesis reduces the great variety of critical phenomena to a small number 
of equivalence classes, the so-called "universality classes", which depend only 
on few fundamental parameters (critical exponents). 
The universality has its origin in the long range character of interactions 
(fluctuations and correlations). 
Close to the transition point, the behavior of the cooperative phenomena 
becomes independent of the microscopic details of the considered system. 
The fundamental parameters determining the universality class are 
the symmetry of the order parameter and the dimensionality of space.

\section{$z$-Scaling}

The  $z$-scaling belongs to the scaling laws with applications not limited 
to the regions near a phase transition. The scaling regularity concerns
hadron production in the high energy proton (antiproton) and nucleus collisions 
(see \cite{22}-\cite{2} and references therein). 
It manifests itself in the fact that the inclusive spectra 
of various types of particles are described with a universal scaling function $\psi(z)$.
The function  $\psi(z)$ depends on a single variable $z$ 
in a wide range of the transverse momentum, 
registration angle, collision energy and centrality. 
The scaling variable has the form
\begin{equation}
z=z_0 \cdot \Omega^{-1}.
\label{eq:1}
\end{equation}
The quantity  $z_0 = {{\sqrt{s_{\bot}}/ [{(dN_{ch}/d\eta|_0)}^cm_N}}]$
is proportional to the transverse kinetic energy $\sqrt {s_{\bot}}$
of a selected binary sub-process responsible for production of the inclusive 
particle with mass $m_a$  and its partner (antiparticle)  with mass $m_b$. 
The multiplicity density $dN_{ch}/d\eta|_0$  
of charged particles in the central interaction region, the nucleon mass $m_N$,  
and the parameter $c$, interpreted as a "specific heat" of the produced medium, completely 
determine the value of $z_0$. 
The quantity $\Omega$ is the maximal relative number of configurations
containing binary sub-processes defined by  
the momentum fractions $x_1$ 
and $x_2$  of colliding hadrons (nuclei), which carry interacting constituents, 
and by the momentum fractions $y_a$ and $y_b$ of objects created directly in these sub-processes,
which carries the inclusive particle and its antiparticle counterpart, respectively.
The relative number of the configurations is given by the function 
\begin{equation}
\Omega = (1-x_1)^{\delta_1}(1-x_2)^{\delta_2}(1-y_a)^{\epsilon_a}(1-y_b)^{\epsilon_b},
\label{eq:2}
\end{equation}
where $\delta_1$  and $\delta_2$  are fractal 
dimensions of the colliding objects, and  $\epsilon_a$  and $\epsilon_b$ 
are fractal dimensions of the fragmentation process in the scattered and recoil direction, respectively. 
The selected binary interaction of the constituents used for 
calculation of the transverse kinetic energy 
$\sqrt {s_{\bot}}$ and $z_0$, 
is defined by the maximum of $\Omega(x_1,x_2,y_a,y_b)$ with the kinematic constraint 
\begin{equation}  
(x_1P_1+x_2P_2-p/y_a)^2=M_X^2. 
\label{eq:3}   
\end{equation}    
The mass  $M_X=x_1M_1+x_2M_2+m_b/y_b$ of the recoil system in the sub-process is 
expressed via momentum fractions and depends implicitly on 4-momenta of the colliding objects and the inclusive particle, $P_1, P_2$  and $p$, respectively.
The constraint (\ref{eq:3}) accounts for the locality of hadron interaction at the constituent
level and sets  a restriction on the momentum fractions  
via kinematics of the constituent sub-process. 
The function $\Omega^{-1}$ represents a resolution
at which a sub-process defined by the fractions $x_1,x_2,y_a,y_b$  can be
singled out of the inclusive reaction.
The scaling variable $z$  has property of a fractal measure. 
It grows in a power-like manner with the increasing resolution $\Omega^{-1}$. 
 
The scaling function $\psi(z)$ is expressed in terms of the experimentally
measured inclusive invariant cross section $Ed^3\sigma/dp^3$, the multiplicity
density $dN/d\eta$ and the total inelastic cross section $\sigma_{in}$ as follows 
\cite{4}
\begin{equation}
\psi(z) = { { \pi } \over { (dN/d\eta)\ \sigma_{in}} } J^{-1} E {
{d^3\sigma} \over {dp^3}}.
\label{eq:4}
\end{equation}
Here $J$ is the Jacobian for the transformation
from $\{ p_T^2, y \}$ to $\{ z, \eta \}$.
The Jacobian depends on kinematic variables
characterizing  the inclusive reaction.
The multiplicity density in the expression (\ref{eq:4})
concerns particular hadrons species. 
The function $\psi(z)$ is normalized to unity
\begin{equation}
\int_{0}^{\infty} \psi(z) dz = 1
\label{eq:5}
\end{equation}
and interpreted as a probability
density to produce an inclusive particle with the corresponding value
of the self-similarity variable $z$.
The flavor independence of $z$-presentation
of inclusive spectra means that the shape of the scaling 
function $\psi(z)$ is the same for hadrons with different 
flavor content over a wide range of $z$ \cite{0,4}.
The scale transformation 
\begin{equation}
z \rightarrow \alpha_F z, \ \ \psi \rightarrow \alpha_F^{-1} \psi
\label{eq:6}
\end{equation}
is used for the comparison of the shapes of the scaling function for different hadron species.
The scale parameter $\alpha_F$ 
depends on type ($F$) of the produced particles.
The transformation  preserves the normalization (\ref{eq:5}) 
and does not destroy the energy, angular, and multiplicity independence
of the $z$-presentation of particle spectra.

\subsection{ Identified hadrons in $p+p$ collisions at RHIC}

Let us remind  the properties of $z$-presentation
of experimental data already found in proton-(anti)proton collisions
at high energies. These are the energy, angular, and multiplicity
independence of scaling function $\psi(z)$ for different types
of hadrons, direct photons, and jets confirmed by numerous
data obtained at U70, ISR, $\rm Sp\bar pS$, Tevatron, and RHIC.
The energy independence of the $z$-presentation of inclusive spectra means
that the shape of the scaling function
is independent on the collision energy $\sqrt s $ over a wide range
of the transverse momentum $p_T$ of produced inclusive particle.
Some results on the energy independence of the $z$-scaling for hadron production in
proton-proton collisions were presented in \cite{0}.
The analyzed data
include negative pions, kaons,
and antiprotons measured at FNAL, ISR, and RHIC energies.
The spectra were measured over a wide transverse
momentum range $p_{T}=0.1-10$~GeV/c.
The cross sections  decrease from $10^2$ to $10^{-10}$~mb/GeV$^2$ in this range.
The strong dependence of the spectra on the collision energy $\sqrt{s}$
increases with transverse momentum.

%***************************************************************
\begin{figure}[h]
\begin{center}
\includegraphics[width=60mm,height=60mm]{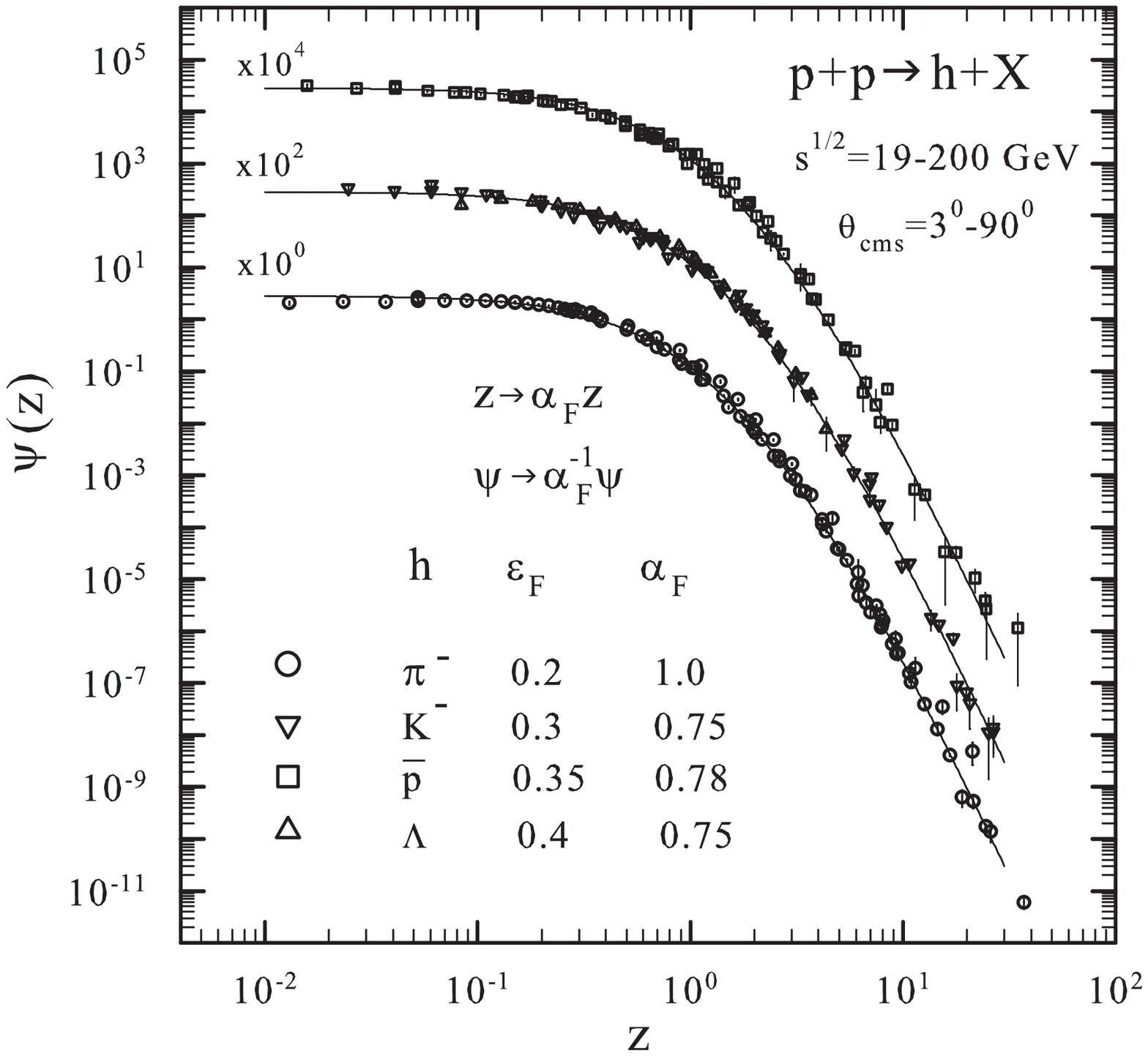}
\hspace{10mm}
\includegraphics[width=60mm,height=60mm]{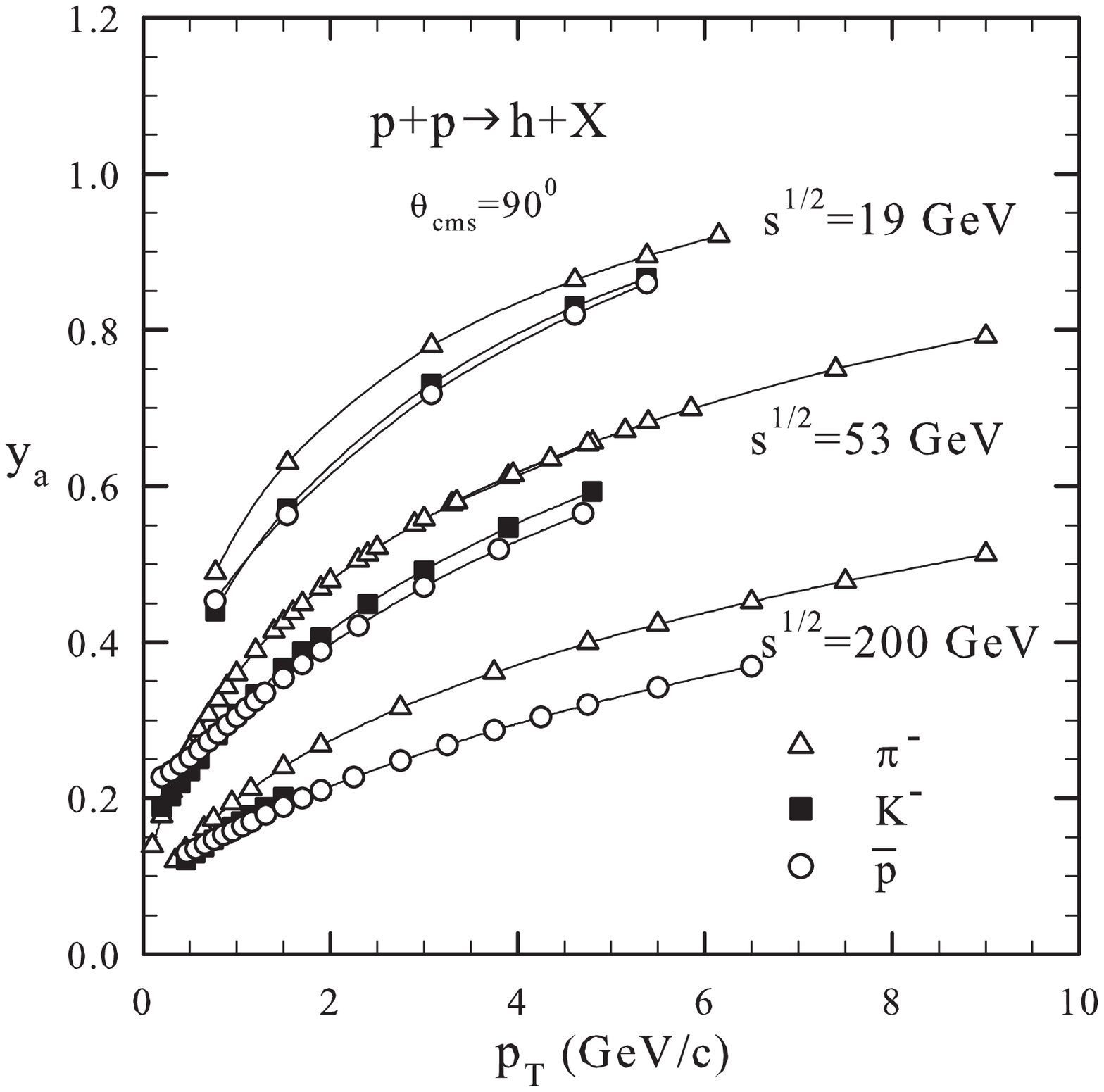}
\vskip 2mm
\hspace*{10mm} (a) \hspace*{72mm} (b)
\begin{minipage}{0.8\textwidth}
\caption{
{\footnotesize
%The flavor independence of $\psi(z)$. 
The inclusive spectra of $\pi^-$, $K^-$,
$\bar p$, and $\Lambda$ 
hadrons produced in $p+p$ collisions in $z$-presentation (a).
Data are taken from
{\protect\cite{CHLM}-\cite{STAR5}}.
The solid lines represent the same curve shifted by multiplicative
factors for reasons of clarity.
The dependence of the fraction~$y_a$  
on the transverse momentum $p_T$ (b)
for  $\pi^-, K^-$, and $\bar p$ produced in the $p+p$ collisions
at $\sqrt s = 19, 53$, and 200~GeV
in the central rapidity region.}
}
\label{f1}
\end{minipage}
\end{center}
\end{figure}
%*************************************************************

Figure \ref{f1}(a) shows the $z$-presentation of
the spectra of $\pi^{-}, K^{-}, \bar p$, 
and $\Lambda's$ produced in $p+p$ collisions over the
range $\sqrt s = 19-200$~GeV and $\theta_{cms}=3^0-90^0$.
The symbols correspond to the data on differential cross sections
measured in the central 
\cite{ISR}-\cite{STAR5}
and fragmentation \cite{CHLM} regions, respectively.
The analysis comprises the inclusive spectra of particles \cite{BSM,BSM2} 
measured up to very small transverse momenta ($p_T\simeq 45$~MeV/c for pions and
$p_T\simeq 120$~MeV/c for kaons or antiprotons).
One can see that the distributions of
different hadrons are sufficiently well described by a single curve over a wide
range of $z=0.01-30$.
The scaling function $\psi(z)$ changes more than ten orders of magnitude.
The solid lines represent the same curve shifted by multiplicative
factors for reasons of clarity.
The same holds for the corresponding data shown with the different symbols.

The $z$-presentation of the transverse momentum distributions 
in proton-proton collisions was obtained for
$\delta_1=\delta_2\equiv\delta$.
We assume that main features of the fragmentation processes
in the scattered and recoil directions can be described by the same
parameter $\epsilon_a=\epsilon_b\equiv\epsilon_F$ which depends
on type $(F)$ of the inclusive particle.
The independence of the scaling function $\psi(z)$ on multiplicity and 
energy was found
for the constant values of the parameters $c=0.25$ and $\delta=0.5$. 
The angular independence of $\psi(z)$ at small angles is sensitive 
to the values of $m_b=m(\pi^+)$,
$m_b=m(K^+)$, and $m_b=m(p)$, for the inclusive production of
$\pi^-$, $K^-$, and antiprotons, respectively. 
The parameter $\epsilon_F$
($\epsilon_{\pi}=0.2, \epsilon_{K}\simeq 0.3, \epsilon_{\bar p}\simeq 0.35,
\epsilon_{\Lambda}\simeq 0.4$) increases with the the mass of the produced hadron.
The indicated values of the parameters
are consistent with the energy,
angular, and multiplicity independence of the $z$-presentation of spectra for 
all types of the analyzed inclusive particles ($\pi, K, \bar p, \Lambda$).
The parameters were found to be independent of kinematic
variables ($\sqrt s, p_T$, and $\theta_{cms}$).
The scale factors $\alpha_F$ are constants
which allow us to describe the $z$-presentation of inclusive spectra
for different hadron species by a single curve.
Based on the obtained results \cite{0} we conclude
that RHIC data on $p+p$ collisions confirm the flavor independence 
of the $z$-scaling including production of particles with very small $p_T$.

The method of construction of the scaling variable $z$ fixes values of the corresponding 
momentum fractions.
The dependence of the fractions $y_a$ and $y_b$ on the kinematic variables
($p_T, \theta_{cms}, \sqrt s $) describes features of the fragmentation processes.
The fraction $y_a$ characterizes dissipation of the energy and momentum of the object
produced by the underlying constituent interaction into the near side of the inclusive particle.
This effectively includes energy loss of the
scattered secondary partons moving in the direction of the registered particle
as well as feed down processes from prompt resonances out of which the inclusive particle
may be created.

Figure \ref{f1}(b) shows the dependence of the fraction $y_a$ on the transverse momentum $p_T$
of $\pi^{-} ,K^{-},\bar p $ particles produced in $p+p$ collisions
at the energy $\sqrt s = 19, 53, 200$~GeV and $\theta_{cms}=90^0$.
All curves demonstrate a non-linear monotonic growth with $p_T$.
It means that the relative energy dissipation associated with the production
of a high $p_T$ particle
is smaller than for the inclusive processes with lower transverse momenta.
This feature is similar for all inclusive reactions at all energies.
The decrease of the fractions $y_a$
with the increasing collision energy is another property of the considered mechanism.
It corresponds to more energy dissipation at higher energies.
This can be due to the larger energy losses and/or due to the heavy prompt resonances.
The third characteristic is a slight decrease of $y_a$ with the mass of the inclusive particle.
It implies more energy dissipation for creation of heavier hadrons
compared to hadrons with smaller masses.

\subsection{Strangeness production in $p+p$ collisions at RHIC}

The strange particles represent a special interest as they contain strange quarks
which are the lightest quarks absent in the net amount in the initial state. 
At the same time, the strange quarks created in the constituent sub-processes are 
substantially heavier than the valence quarks in the colliding protons. 
The self-similarity of such interactions, expressed by the same form of the scaling function,
results in different properties of the constituent collisions and fragmentation processes 
as compared to those which underlay the production of the non-strange particles.
The scaling behavior of $\psi(z)$ for strange particles 
could give more evidence in support of unique description of $p+p$ interaction 
at a constituent level 
and could provide a good basis for study of peculiarities  of the strangeness production
in nuclear collisions, as well as for study of the origin of the strangeness itself.

The self-similarity of strange hadron production  
was studied \cite{4} using data \cite{KS0_PX}-\cite{K*_PX} on inclusive cross-sections 
of $K_S^0, K^-$, $K^{*0}, \phi $ 
mesons  measured in proton collisions at RHIC. 
The data on strange particle spectra 
\cite{ISR}, \cite{KS0_44}-\cite{K*_NA49}
obtained by the BS, CCRS, CDHW, AFS, NA61/SHINE, and NA49 Collaborations 
were used in the analysis as well.

\begin{figure}[t]
\begin{center}
\includegraphics[width=60mm,height=60mm]{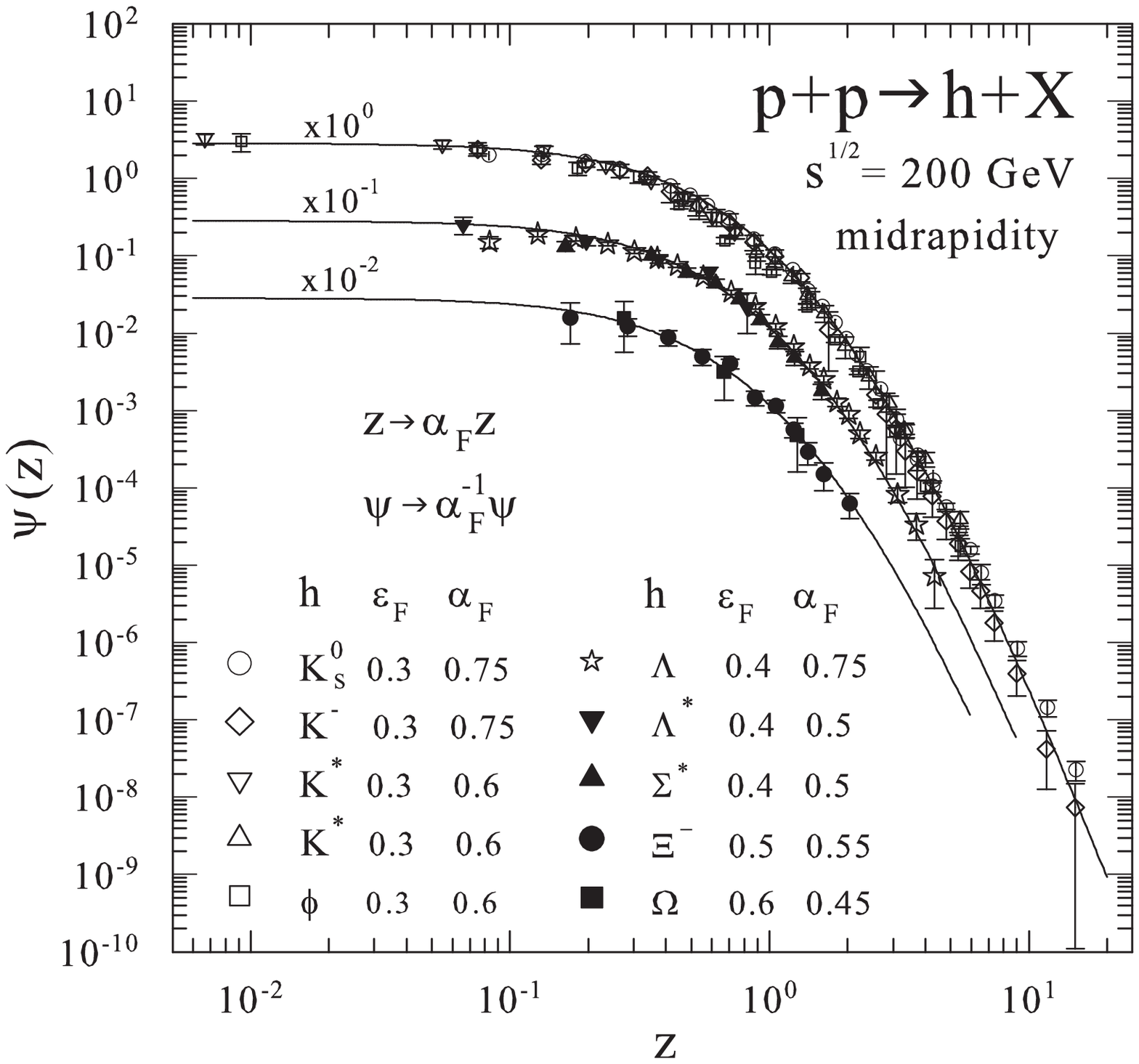}
\hspace{10mm}
\includegraphics[width=60mm,height=60mm]{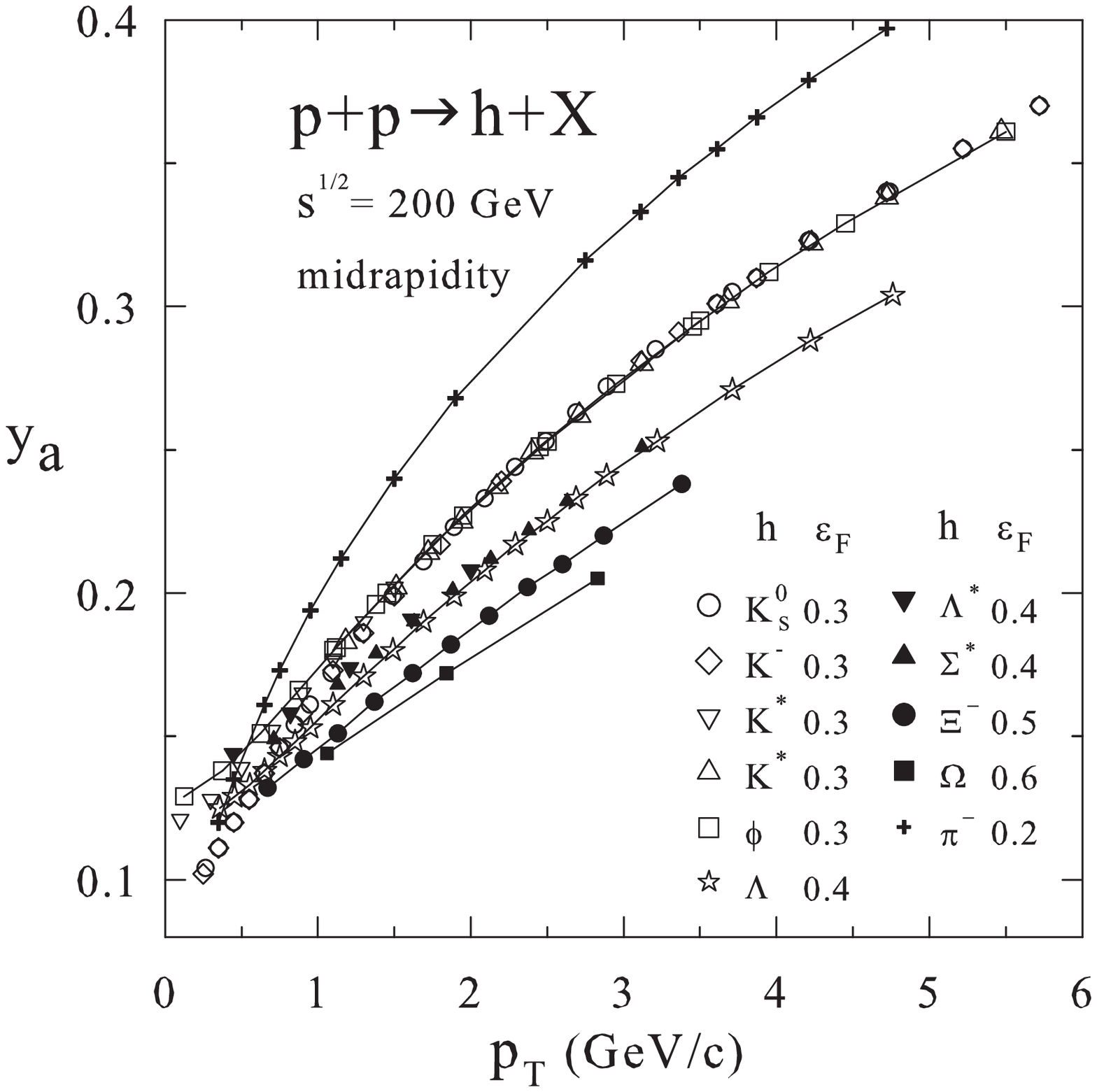}
\vskip 2mm
\hspace*{10mm} (a) \hspace*{72mm} (b)
\begin{minipage}{0.8\textwidth}
\caption{
{\footnotesize
Scaling function $\psi(z)$ (a) and momentum fraction $y_a$ (b) 
for strange ($K_S^0, K^-$, $K^{*0}, \phi $, $\Lambda, \Lambda^{*} $,
$\Sigma^{*} , \Xi^{-}, \Omega$) hadrons produced 
in $p+p$ collisions at $\sqrt s =200$~GeV 
and  $\theta_{cms}\simeq 90^{0}$. 
Experimental data are taken from
\cite{STAR5}-\cite{K*_PX}.
The solid line in (a) is fit of $\psi(z)$ for $\pi^{-}$
mesons \cite{Pim_ST}.
The points in (a) and (b)  are calculated for $\delta=0.5$ and for the 
indicated values of $\epsilon_F$ and $\alpha_F$.}
}
\label{f2}
\end{minipage}
\end{center}
\end{figure}

Figure \ref{f2}(a)  shows $z$-presentation  of the transverse momentum 
spectra \cite{STAR5}-\cite{K*_PX} of strange mesons and baryons measured 
in $p+p$ collisions at the energy $\sqrt s = 200$~GeV 
in the central rapidity region at RHIC.
The symbols representing data on differential inclusive cross sections
include baryons which consist of one, two and three strange
valence quarks. 
The multiplicative factors $10^0, 10^{-1}$, and $10^{-2}$ are used 
to show  the data $z$-presentation separately  for mesons, 
single-strange ($\Lambda, \Lambda^*, \Sigma^* $) and multi-strange 
($\Xi^-, \Omega$) baryons, respectively.
The symbols for different particles are shown for 
the indicated values of the parameters $\epsilon_{F}$ and~$\alpha_{F}$. 
They are reasonably well described by the solid curve representing
a reference line ($\alpha_{\pi}=~1$) for $\pi^{-}$ mesons 
obtained from analysis \cite{0} of pion spectra.
It is consistent with the energy, angular and multiplicity
independence of the scaling function for different hadrons.
The fragmentation dimension
$ \epsilon_F $ for strange mesons is larger than for pions ($\epsilon_{\pi}=0.20\pm 0.01$).
It suggests larger energy loss by production of mesons with
strangeness content.  The fragmentation dimension 
for strange baryons grows with the number of the 
strange valence quarks. 

%**************************************************************
%******************************************************************************

Figure \ref{f2}(b) demonstrates the $p_T$-dependence of the momentum fraction $y_a$ 
of strange hadrons and $\pi^-$ mesons 
produced in $p+p$ collisions  at $\sqrt s = $~200~GeV.  
The fraction $y_a$ increases with the transverse momentum for all particles.  
The  energy loss $ \Delta E_q/E_q = (1-y_a)$ depends on value
of the fragmentation dimension $\epsilon_F$. 
As one can see, the relative energy loss decreases 
with the increasing $p_T$ for all particles.
For a given $ p_T > 1$~GeV/c, 
the energy loss is larger for strange baryons than 
for strange mesons. The growth indicates increasing 
tendency with larger number of strange valence quarks inside the strange baryon,   
$(\Delta E/E)_{\Omega}>(\Delta E/E)_{\Xi^-}>(\Delta E/E)_{\Lambda}\simeq (\Delta E/E)_{\Lambda^*}\simeq(\Delta E/E)_{\Sigma^*}$.

\subsection{Top-quark production at LHC and Tevatron}

The top quark is the heaviest known elementary particle. 
It was discovered at the
Tevatron in 1995 by the CDF and
D\O \ Collaborations 
\cite{CDFtop, D0top} 
at a mass of around 170~GeV.
The first measurements of the differential cross section
as a function of the transverse momentum of the top quark
were presented by the D\O \ Collaboration \cite{D0_top1}.
It is expected that $top$ physics is extremely important 
in the search for new and for the study of known symmetries in high-$p_T$ region.

%******************************************************************
\begin{figure}[h]
\begin{center}
\vskip 5mm
\hspace*{0mm}
\includegraphics[width=60mm,height=60mm]{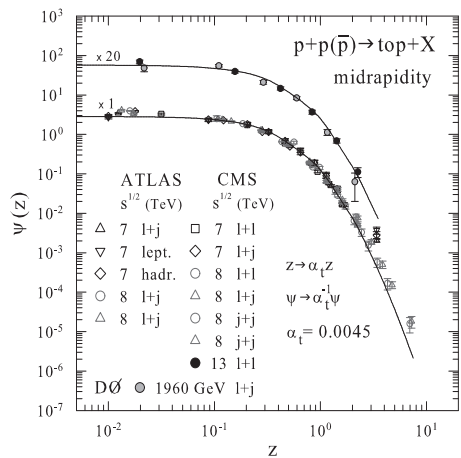}
\vskip -0mm
\begin{minipage}{0.8\textwidth}
\caption{
{\footnotesize
The scaling function $\psi(z)$ of the top-quark production in $p+p$ 
and $p+\bar p$ collisions at the LHC energies 
$\sqrt s =7,8, 13$~TeV and at the Tevatron energy $\sqrt s =1.96$~TeV.
The symbols 
denote the experimental data obtained 
by the CMS \cite{CMS_top7}-\cite{CMS_top13}, 
ATLAS \cite{ATLAS_top7}-\cite{ATLAS_top8_1} 
and  
D\O \ 
\cite{D0_top2}
Collaborations. 
The solid line is a reference curve corresponding to $\pi^-$-meson production 
in $p\!+\!p$ collisions.} 
}
\label{f3}
\end{minipage}
\end{center}
\end{figure}
%****************************************************************************

Figure \ref{f3} shows the $z$-presentation \cite{2} of the spectra of top-quark production 
obtained in $p+p$ collisions at the LHC energies $\sqrt s = 7,8$, and 13~TeV 
in the central rapidity region. 
The measurements of the inclusive cross sections were performed 
by the CMS \cite{CMS_top7}-\cite{CMS_top13} and 
ATLAS \cite{ATLAS_top7}-\cite{ATLAS_top8_1} Collaborations 
in the dilepton and jet channels. 
The data include measurements over a wide range of the transverse momentum 
$30<p_T<1000$~GeV/c.
The $z$-scaling of $\pi^{-}$-meson spectra 
shown by the solid line serves as a reference curve.
The values of the fractal dimension $\delta=0.5$ and the parameter $c=0.25$
are the same as used in our previous analyses \cite{0,4} for other hadrons.
We have set $\epsilon_{top}=0$, 
as negligible energy loss 
is assumed in the elementary $t\bar{t}$ production process. 
The scale parameter $\alpha_F$ in the transformation (\ref{eq:6}) 
is found to be $\alpha_{top}\simeq 0.0045$. 
The data on the top-quark production \cite{D0_top2}
in $\bar p+p$  collisions obtained by the D\O \
Collaboration at the Tevatron energy $\sqrt s=1.96$~TeV 
are compatible with the LHC data in $z$-presentation. 
The scaling function $\psi(z)$
demonstrates energy independence over a wide range 
of the self-similarity parameter $z$.

Based on the above comparison we conclude
that the LHC and Tevatron data on  inclusive spectra of the top quark    
support the flavor independence
of the scaling function $\psi(z)$ over the interval of $z=0.01-8$. 
This result gives us indication on the self-similarity of top-quark production
in $p+p$ and $\bar p+p$ interactions up to the top-quark transverse momentum $p_T = 1$~TeV/c 
and for a wide range of the collision energy $\sqrt s = 1.96, 7, 8$ and 13~TeV.

\subsection{Jet production at LHC and Tevatron}

Jets are traditionally considered as best
probes of constituent interactions at high energies.
They are of interest both for study of jet properties itself 
and in search for new particles identified by the jets.
In hadron collisions, jet is a direct evidence of hard interaction 
of hadron constituents (quarks and gluons).   
The data on inclusive cross sections of
jet production in $p+p$ collisions at the LHC 
energies $\sqrt s =2760,7000$, and  8000~GeV
\cite{ATLAS_ja2760}-\cite{ALICE_j2760}
were analysed \cite{2} in the framework of  $z$-scaling.
We used the parameter
values $c=1$, $\delta=~1$, $\epsilon_{jet}=0$, and $m_a=m_b=0$ for the analysis.
The results are compared with $z$-presentation 
of jet spectra in 
$\bar p+p$ collisions at the Tevatron energies
$\sqrt s = 630, 1800, 1960$~GeV 
\cite{Abbott}-\cite{CDF_ja1960}.

%********************************************
\begin{figure}[h]
\begin{center}
%\vskip 0.4cm
\hspace*{0mm}
\includegraphics[width=60mm,height=60mm]{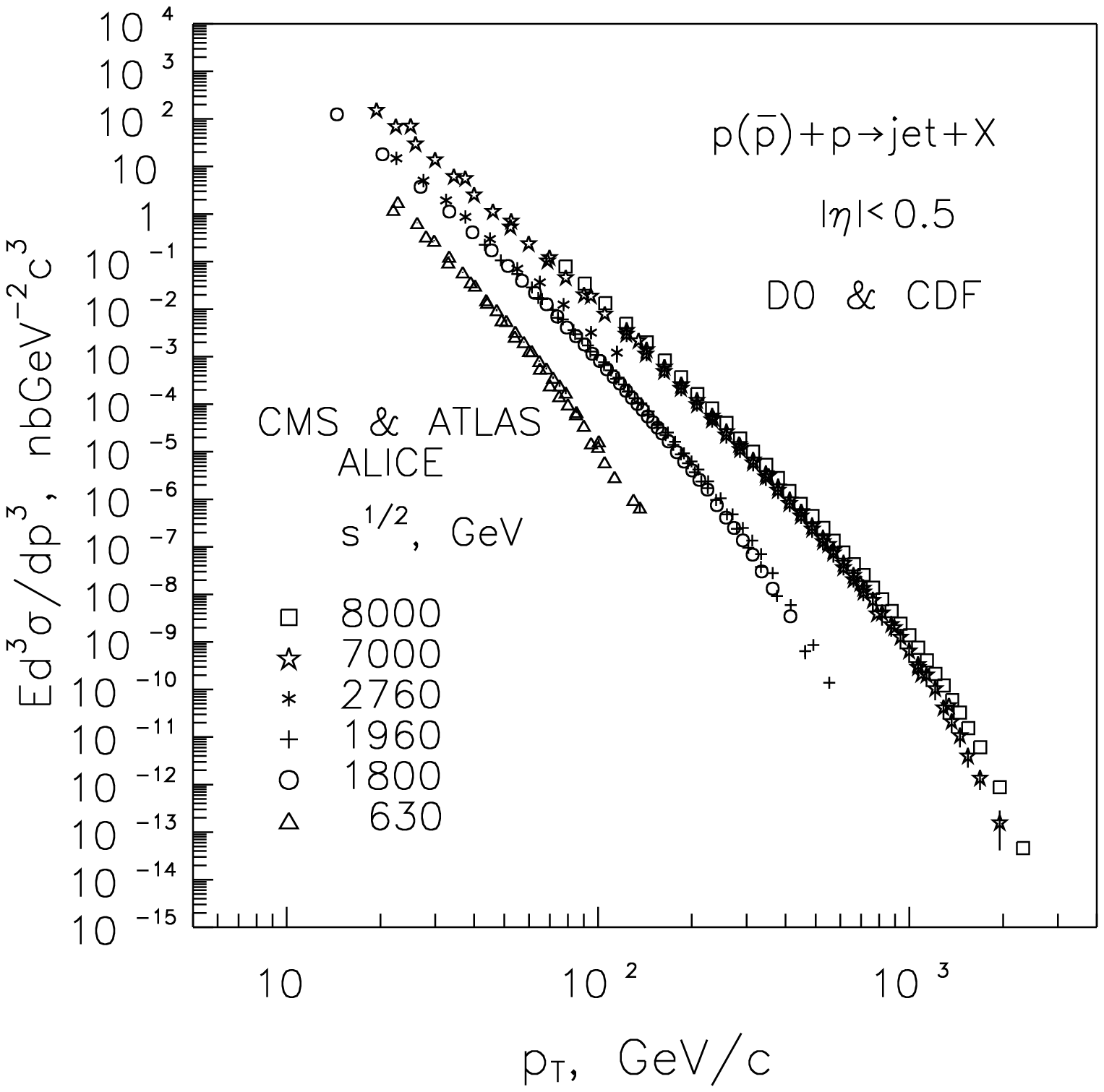}
%\vskip -5mm
\hspace{4.mm}
\includegraphics[width=60mm,height=60mm]{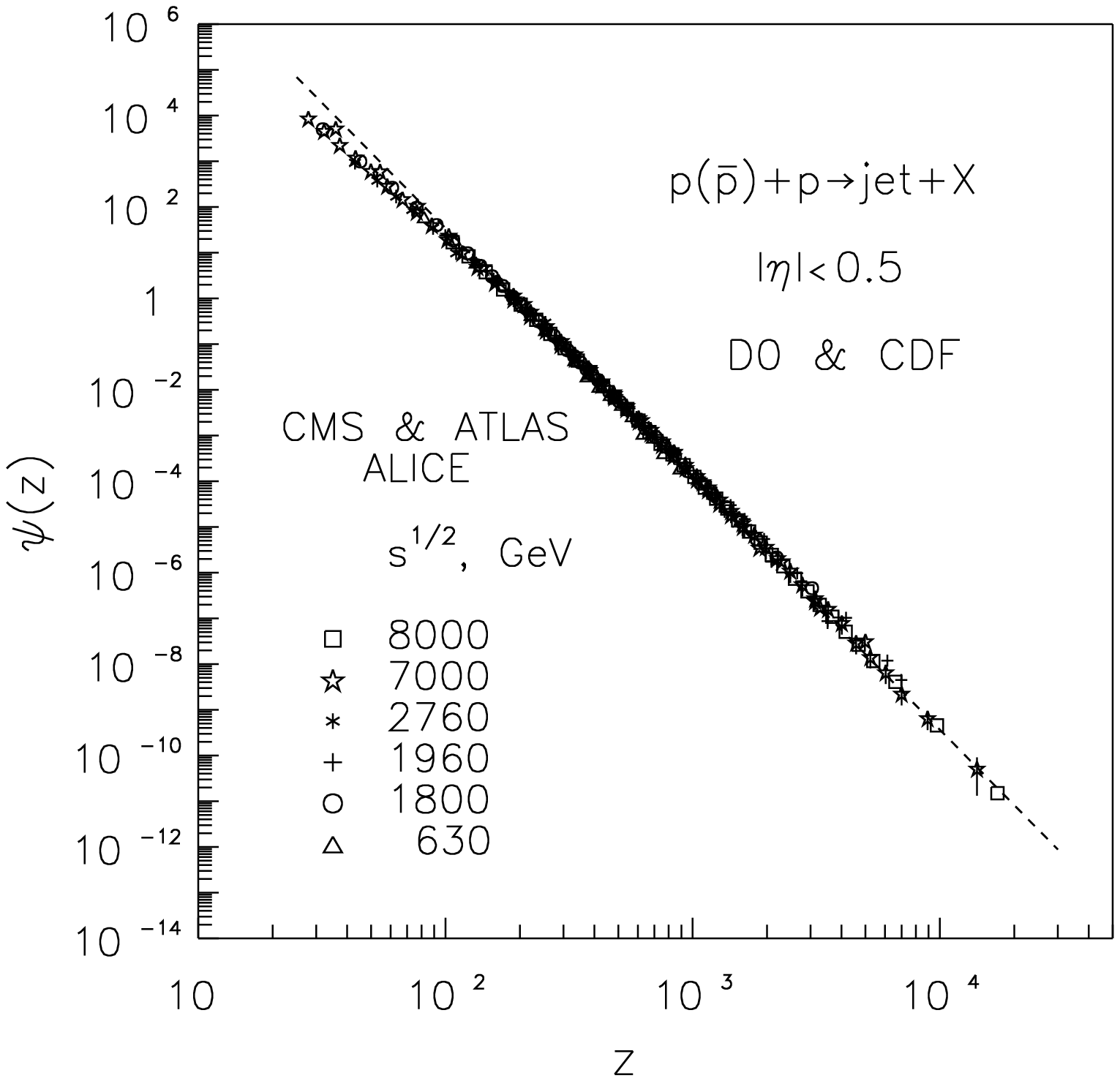}
\vskip 2mm
\hspace{10mm}    (a) \hspace*{60mm} (b)
\begin{minipage}{0.8\textwidth}
\caption{
{\footnotesize
The inclusive spectra of jet production in $p+p$ and $\bar p+p$ 
collisions in $p_T$ (a)
and  $z$-presentation (b) measured at $\theta \simeq 90^0$.
The symbols denote the ATLAS 
\cite{ATLAS_ja2760,ATLAS_ja7000,ATLAS_ja8000}, CMS \cite{CMS_ja8000,CMS_ja7000} and 
ALICE \cite{ALICE_j2760} 
data obtained in $p+p$ collisions at 
$\sqrt s = 2760,7000,8000$~GeV, 
and the  D\O \ \cite{Abbott}-\cite{D0_ja1960} 
and CDF \cite{Abe}-\cite{CDF_ja1960} data 
obtained  in $\bar p+p$ collisions at $\sqrt s = 630, 1800, 1960$~GeV. }
}
\label{f4}
\end{minipage}
\end{center}
\end{figure}
%******************************************************

Figure \ref{f4}(a) shows  $p_T$-dependence of jet spectra measured 
in the central pseudorapidity window $|\eta|<0.5$ by the  
ATLAS 
\cite{ATLAS_ja2760,ATLAS_ja7000,ATLAS_ja8000}, 
CMS 
\cite{CMS_ja8000,CMS_ja7000}, 
and ALICE 
\cite{ALICE_j2760}
Collaborations 
at the LHC and the spectra obtained in the mid-rapidity region by the 
D\O \ 
% \ \cite{Abbott,Abbott1,Abbott2,Abbott3,Begel,D0_ja1960}
\ \cite{Abbott}-\cite{D0_ja1960}
and CDF \cite{Abe}-\cite{CDF_ja1960} 
Collaborations at the Tevatron.
The data collected by the CMS Collaboration at $\sqrt s = 8000$~GeV 
correspond to the integrated luminosity of 19.7~fb$^{-1}$.
The spectra were measured up to the transverse momentum $p_T = 2500$~GeV/c.
The measurement based on the
data collected with the ATLAS detector at $\sqrt s = 8000$~GeV corresponds to
an integrated luminosity of 20.2~fb$^{-1}$.
The ATLAS data cover the range $ 70<p_T < 2500$~GeV/c.
The distributions change by many orders of magnitude within the analyzed $p_T$-range.
As can be seen from Fig. \ref{f4}(a), the strong dependence of the spectra on the collision
energy  $\sqrt s$ increases with the transverse momentum.

Figure \ref{f4}(b) demonstrates 
the energy independence of $\psi(z)$ for jet production and its power behavior 
over the range $\sqrt s = 630-8000$~GeV at $\eta \simeq 0$.
The scaling function changes more than twelve orders of magnitude
and can be described by a power law $\psi(z)\sim z^{-\beta}$
over a wide $z$-range. 
The slope parameter $\beta$ is energy independent in the large region of $\sqrt s$.
The dashed line 
corresponds to the asymptotic behavior of $\psi(z)$.  
The data obtained at the LHC confirm results of the analysis \cite{5}
of jet spectra measured by the D\O \ and CDF Collaborations 
in $\bar p+p$ collisions at the Tevatron with parameters $\delta=1$ and $c=1$.

\subsection{BES-I program in $Au+Au$ collisions at RHIC}

Experimental results from RHIC and LHC  support the
hypothesis that a new state of nuclear matter is created  
in the collisions of heavy ions at high energy.
The created matter with quark and gluon degrees of freedom, 
the Quark-Gluon Plasma, reveals features of a strongly-coupled medium.
It is assumed that the medium produced
in heavy-ion collisions is thermalized.
The phase diagram of nuclear matter is usually presented in the temperature-baryon 
chemical potential plane $\{T, \mu_B$\}. Both quantities
can be varied by changing the energy and centrality of the nuclear collisions 
and the type and momentum of the produced particles.
The idea of the Beam Energy Scan (BES) program at RHIC  
is to scan 
the phase diagram of nuclear matter from the top RHIC energy 
to the lowest possible energy achievable on this collider 
and compare it with the phase diagram predicted by QCD theory
\cite{BESI1,BESI6}.
The program is aimed 
to perform systematic investigation and data analysis 
of particle production in the heavy-ion interactions  
over a wide range of collision energy 
and centrality. 
The systematic measurements performed 
with heavy-ions are of great interest to search for critical phenomena 
in a broad range of kinematic variables.

We extend the applicability of the self-similarity principle 
to the description of hadron production in nucleus-nucleus collisions.
The self-similarity concerns fractal structure of the colliding
objects, interaction of their constituents and fractal character of fragmentation 
processes in the final state.
This physical principle is assumed to be valid also in 
the high-density and high-temperature phase
in which quark and gluon degrees of freedom dominate.

%****************************************
\begin{figure}[h!]
\begin{center}
\vskip 4mm
\hspace*{0mm}
\includegraphics[width=60mm,height=60mm]{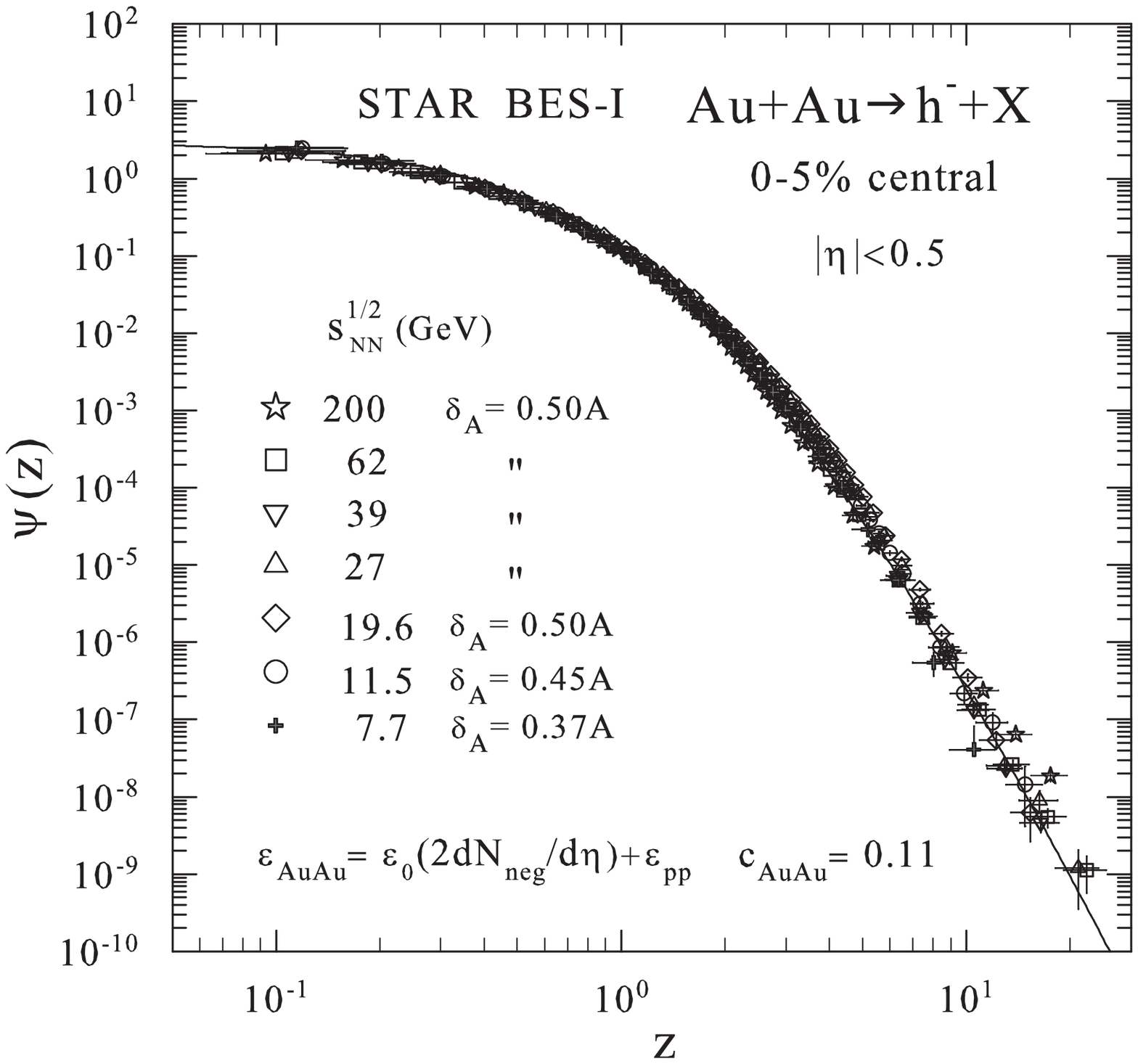}
\hspace{4.mm}
\includegraphics[width=60mm,height=60mm]{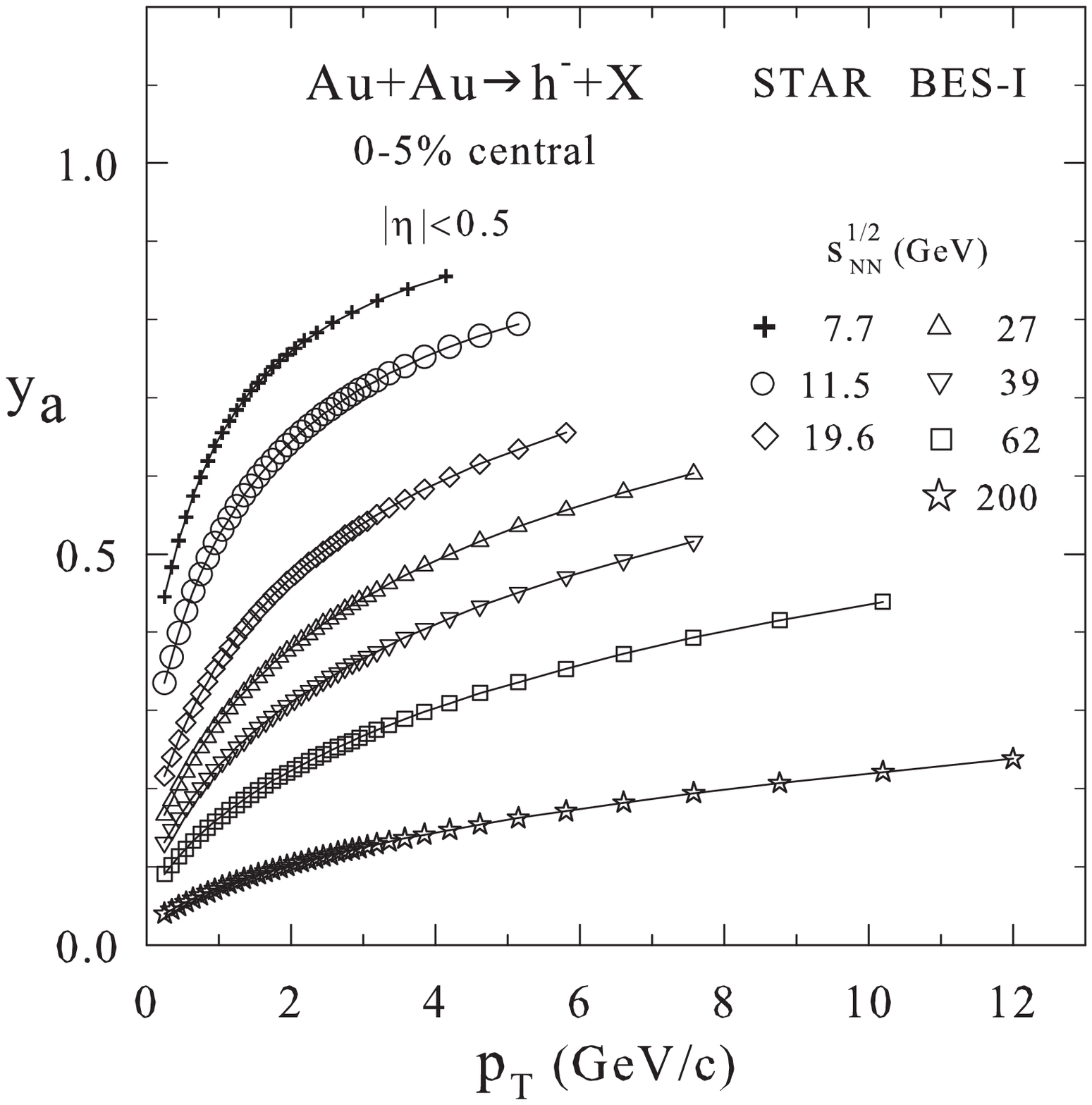}
\vskip 0mm
\hspace{10mm}    (a) \hspace*{60mm} (b)
\begin{minipage}{0.8\textwidth}
\caption{
{\footnotesize
Scaling function $\psi(z)$ (a) and the momentum fraction $y_a$ 
in dependence on the transverse momentum $p_T$ (b)
for negative hadrons produced in 
$(0-5)\%$ central $Au+Au$ collisions 
at  $\sqrt {s_{NN}} =$ 7.7, 11.5, 19.6, 27, 39, 62.4 and 200~GeV
\protect\cite{1}.
The symbols correspond to experimental data \protect\cite{BESI7}
measured by the STAR Collaboration at RHIC.
}
}
\label{f5}
\end{minipage}
\end{center}
\end{figure}
% ********************************************************************

Figure \ref{f5}(a) shows the scaling function $\psi(z)$ for negative hadrons \cite{1} 
produced in the $(0\!-\!5)\%$  central 
$Au\!+\!Au$ collisions at different energies $\sqrt {s_{NN}}=7.7-200$~GeV.
The symbols correspond to the spectra \cite{BESI7} measured
by the STAR Collaboration in the pseudorapidity window $|\eta|<0.5$.
The $z$-presentation of the spectra
demonstrates energy independence of the function $\psi(z)$ over the analyzed kinematic range. 
Moreover, the symbols representing the nuclear data 
are in reasonable agreement   
with the solid curve which depicts the $z$-scaling 
of $h^-$ particles produced in $p+p$ collisions over the range $\sqrt{s}=11.5-200$~GeV.
The same energy independence of $\psi(z)$ is valid \cite{1} for different centrality classes  
of $Au\!+\!Au$ collisions. 

The scaling was obtained for the multiplicity dependent fragmentation dimension in the form 
$\epsilon_{AuAu}=\epsilon_0(2dN^{AuAu}_{neg}/d\eta)+\epsilon_{pp}$ 
with a suitable choice of $\epsilon_0$,
and for the constant values of the model parameters $c_{AuAu}=0.11$, $\delta_A=A\delta$, 
$\delta=0.5$, and $\epsilon_{pp}=0.2$ 
at $\sqrt{s_{NN}}\gtrsim~19.6$~GeV.
The parameter $\epsilon_{0}$ shows logarithmic increase with $\sqrt {s_{NN}}$ \cite{1}.
It reflects growing tendency of the suppression of hadron yields 
in the central collisions of heavy nuclei as the collision energy and multiplicity increases.
A decrease of the parameters 
$\delta_{Au}$, $\delta$, and $\epsilon_{pp}$ with energy for $\sqrt{s_{NN}}<~19.6$ ~GeV
indicates smearing of the manifestations of fractality of the interacting objects
and fragmentation process at low energies.

The $z$-scaling  of negative hadron production 
in $Au+Au$ collisions was found in the environment with different multiplicity densities.  
The multiplicity scan of particle yields in nucleus-nucleus collisions at different energies gives 
complementary information on the production mechanisms in nuclear medium.    
The scaling can be interpreted as a result of a self-similar modification 
of elementary sub-processes by the created medium. 
We assume that verification of the scaling behavior of $\psi(z)$ 
at even higher (lower) $z$ at high multiplicities could give new restrictions
on the model parameters. 
Based on the self-similarity arguments, one can search for changes of the 
scaling parameters in the nuclear systems relative to ones established 
in $p+p$ interactions. 
A discontinuity or abrupt change of the structural ($\delta_A $) and 
fragmentation ($\epsilon_{AA}$) fractal dimensions and the
"specific heat" ($c_{AA}$) may indicate to a signature of a phase transition 
or a critical point in the matter produced in nuclei collisions.
Therefore quest for irregularities in the behavior of the 
$z$-scaling parameters is inspired by searching of the location
of a phase boundary and a critical point, 
which is of great interest. 
 
The increase of $\epsilon_{AuAu}$ with multiplicity density
is connected with a decrease of the momentum fraction $y_a$, 
representing larger energy loss in final state for large centralities (multiplicities).
The value of $y_a$ is a characteristic which describes relative energy 
dissipation $\Delta E_q/E_q=(1-y_a)$ in the final 
state by production of an inclusive particle. 
The energy losses depend on the
traversed medium which converts them into the multiplicity of the
produced particles.
This leads to the fact that the more produced multiplicity 
$ N^{AA}_{ch}\simeq 2N^{AA}_{neg}$ per unit of (pseudo)rapidity, the larger  
energy loss of the secondary particles.
The multiplicity density
characterizes the produced medium and is connected in this way to
the energy loss in this medium.

The amount of the relative energy loss, expressed by $y_a$ and its dependence 
on $\sqrt{s_{NN}}$, multiplicity, and transverse momentum of produced particle, 
has relevance to the evolution of the matter created in nuclei collisions.
The energy and multiplicity characteristics of the energy loss
can be sensitive to the nature of the medium and can
reflect changes in the fragmentation process which may occur 
by production of inclusive particles.
Figure \ref{f5}(b) shows the dependence of the fraction $y_a$ 
on the  transverse momentum $p_T$ for $h^-$ hadrons produced 
in the $(0-5)\%$  central $Au+Au$ collisions for different energies. 
A monotonic growth of $y_a$ with the momentum $p_T$ is found for all energies.
This means that the relative energy dissipation associated with the production of  
a high-$p_T$ particle is smaller than for the inclusive
process with lower transverse momenta.
At given $p_T$, the decrease of $y_a$ with $\sqrt{s_{NN}}$ shows 
considerable growth of the relative energy loss, as the collision energy increases.

\section{$z$-Scaling and entropy}

The parameters used in the $z$-scaling scheme can be interpreted in terms of some
thermodynamic quantities (entropy, specific heat, chemical potential) of a multiple particle system.
There exists a profound connection between the variable $z$ and entropy \cite{22,0}.
The scaling variable is proportional to the ratio
\begin{equation}
z \sim \frac {\sqrt {s_{\bot}}}{W}
\label{eq:7}
\end{equation}
of the transverse kinetic energy $\sqrt {s_{\bot}}$
and the maximal value of the function
\begin{equation}
W(x_1,x_2,y_a,y_b) = (dN_{ch}/d\eta|_0)^c \cdot \Omega(x_1,x_2,y_a,y_b) 
\label{eq:8}
\end{equation}
in the space of the momentum fractions, constrained by the condition (\ref{eq:3}).
The function $W$ is proportional to the number of all parton and hadron configurations 
of the colliding system which contain the constituent configuration defined 
by particular values of the momentum fractions $x_1$, $x_2$, $y_a$ and $y_b$.

According to statistical physics, entropy of a system is given by a number $W_S$
of its statistical states as follows
\begin{equation}
S = \ln W_S.
\label{eq:9}
\end{equation} 
The most likely configuration of the system is given by the maximal value of $S$. 
For the inclusive reactions, the quantity $W_S$ is the number of all parton and hadron configurations
in the initial and final state of the colliding system which can contribute
to the production of inclusive particle. The configurations comprise all constituent
configurations that are mutually connected by independent sub-processes. The binary
sub-processes corresponding to the production of the inclusive particle with the 
4-momentum $p$ are subject to the condition (\ref{eq:3}). The underlying sub-process, which
defines the variable $z$, is singled out from these sub-processes by the
principle of maximal entropy $S$. The absolute number of the configurations,
$W_S = W\cdot W_0 $, is given up to a multiplicative constant $W_0$. 
Its value is restricted by the positiveness
of entropy above some scale characterized by a maximal resolution $(\Omega^{-1})_{max}$.
For the infinite resolution at fractal limit, $W_0$ is classically infinity.
Denoting $S_0=\ln W_0 $ and using relations (\ref{eq:2}), (\ref{eq:8}) and(\ref{eq:9}), 
we write the entropy of system of the considered configurations as follows
\begin{equation}
S=c\cdot \ln (dN_{ch}/d\eta|_0) +
\ln(1-x_1)^{\delta_1}(1-x_2)^{\delta_2}(1-y_a)^{\epsilon_a}(1-y_b)^{\epsilon_b}+S_0.
\label{eq:10}
\end{equation}
In thermodynamics, the entropy for an ideal gas is given by the formula
\begin{equation}
S = c_V\cdot \ln T + R\cdot \ln V + S_0.
\label{eq:11}
\end{equation} 
 Here $R$ is an universal constant. 
The specific heat $c_V$, temperature $T$ and volume $V$ characterize a state of the system. 

There is an analogy between expressions (\ref{eq:10}) and (\ref{eq:11}).
The analogy is supported
by the plausible idea that interactions of the extended objects like hadrons and
nuclei can be treated at sufficiently high energies as a set of independent collisions
of their constituents. 
Such concept justifies a division of the system into the part
comprising a binary sub-process which can contribute to production of the inclusive
particle with 4-momentum $p$, and the remaining part  containing all other 
microscopic configurations which lead to the produced multiplicity. 
Maximization of the entropy (\ref{eq:10}), 
constrained by the condition (\ref{eq:3}), corresponds to the maximal
entropy of the remaining part of the system. 
The multiplicity density of  particles produced 
in the central interaction region characterizes a "temperature" created 
in the system.
Provided the system is in a local equilibrium, there exists a simple relation
$dN_{ch}/d\eta|_0\sim T^3$ for high temperatures and small chemical potentials. 
Using the mentioned analogy, the model parameter $c$ plays a role of a "specific heat" 
of the produced matter. 
The second term in (\ref{eq:10}) depends on the volume of the rest of the system
in the space of the momentum fractions $\{x_1, x_2, y_a, y_b\}$. 
The volume is a product of the complements of the fractions with exponents which are generally fractional
numbers, $V=l_1^{\delta_1}\cdot l_2^{\delta_2}\cdot l_a^{\epsilon_a}\cdot l_b^{\epsilon_b}$.
This analogy emphasizes once more the interpretation
of the model parameters $\delta_1, \delta_2, \epsilon_a$ and $ \epsilon_b$ as fractal dimensions.

The entropy (\ref{eq:10}) increases
with the multiplicity density $dN_{ch}/d\eta|_0$ and 
decreases with the increasing resolution $\Omega^{-1}$.
The minimal resolution with respect to all binary
sub-processes satisfying the condition (\ref{eq:3}), 
which singles out the corresponding sub-process, 
is equivalent to the principle of the maximal
entropy $S$ of the rest of the system.

\subsection{Maximum entropy principle and conservation of fractal cumulativity}

According to the assumption of fractal self-similarity of hadron structure and
fragmentation process and due to the locality of binary interactions of hadron
constituents, there exists a conservation law of a scale dependent quantity
characterizing hadron interactions at a constituent level.
The quantification of such a statement is based on the maximum entropy principle.
The conservation law reflects a symmetry of transformation 
of one fractal structure into another one at all scales.
The corresponding symmetry is
encoded in the functional form of the fractions ${x_1, x_2, y_a, y_b}$ which follows
from the requirement of the maximal entropy~(\ref{eq:10}).
The conditions for the maximization of the entropy with the constraint (\ref{eq:3}) 
determine specific dependences of the fractions 
on the kinematics of the inclusive reaction.
As shown in \cite{3}, the momentum fractions satisfy the following
equality
\begin{equation}
\delta_1\frac{x_1}{1-x_1}+\delta_2\frac{x_2}{1-x_2} =
\epsilon_a\frac{y_a}{1-y_a} + \epsilon_b\frac{y_b}{1-y_b}.
\label{eq:12}
\end{equation}
This equation represents a conservation law for the quantity
\begin{equation}
C(D,\zeta) = D\cdot g(\zeta ), \ \ \ g(\zeta) = \frac {\zeta}{1-\zeta}.
\label{eq:13}
\end{equation}
The symbol $D$ means a fractal dimension and $\zeta$ 
is the corresponding momentum fraction. 
The conservation law (\ref{eq:12}) holds for any
inclusive reaction with arbitrary momenta $P_1, P_2$ 
and $p$ of the colliding and inclusive particles
which determine corresponding level of resolution.
We name the quantity $C(D,\zeta )$ as the "fractal cumulativity" of a fractal-like structure   
with the dimension $D$ carried by its constituent 
with the momentum fraction $\zeta$. 
The conservation law for this quantity is formulated as follows: 

{\it The fractal cumulativity  before a constituent interaction is equal to the fractal cumulativity
after the constituent interaction for any binary constituent sub-process,}

\begin{equation}
\sum_i^{in} C(D_i,\zeta_i) = \sum_j^{out} C(D_j,\zeta_j).
\label{eq:14}
\end{equation}

The quantity $C(D,\zeta)$ characterizes
the property of a fractal-like object or a fractal-like process with the dimension $D$ 
to form a structural aggregate 
with certain degree of local compactness, which carries its momentum fraction $\zeta$.
The value of the fractal cumulativity  is a measure of the ability of the fractal systems 
to create the aggregated sub-structures.
This cumulative feature of the internal structure of hadrons and nuclei is connected 
with formation of hadron constituents interacting in the underlying  
sub-processes locally. 
It is in agreement with the Heisenberg uncertainty principle.
The aggregation property in the final state concerns  formation of the
produced particles in the fractal-like fragmentation process.

The scale dependence of the conserved quantity $C(D,\zeta)$ is given by the function $g(\zeta)$. 
Due to the general way of its derivation, 
the form of $g(\zeta)$ is the same for arbitrary fractal objects
(different hadrons, nuclei, hadron constituents, jets,
guarks, gluons, etc.) participating in the high energy interactions. 
The dimension $D$ is considered to be a new and unique characteristic 
of the related fractal structures such as mass, charge and spin. 
The fractal cumulativity corresponding to different momentum fractions satisfy  
the following relations
\begin{equation}
C(D,\zeta^{"}) = C(D,\zeta)+C(D,\zeta^{'})+D^{-1}\cdot C(D,\zeta)\cdot C(D,\zeta^{'}), 
\label{eq:15}
\end{equation}
\begin{equation}
\zeta^{"} = \zeta+\zeta^{'}-\zeta \cdot \zeta^{'}.
\label{eq:16}
\end{equation}
This is a composition rule connecting the values of the cumulativity $C(D,\zeta)$ of
the same fractal structure at different levels of its aggregation.

We would like to note that (\ref{eq:12}) was
derived by the formulae expressed in terms of the Lorentz invariant quantities. It
means that the conservation law for the fractal cumulativity holds in any motion inertial
frame in the unchanged form. 
The cumulativity $C(D,\zeta)$ is therefore a relativistic
invariant with respect to motion. 
The quantity manifests itself in hadron interactions
at a constituent level. 
Large values of the fractal cumulativity of the interacting
hadron structures can be obtained by  increasing the resolution or compactness of the system. 
The state of resolution revealed in measurements of the inclusive particles depends on fractal
dimensions of the interacting objects and fractal dimensions of the fragmentation
processes in the final state.

\subsection{Entropy decomposition and quantization of fractal dimensions}
 
The fractality of hadron structure and fragmentation process manifests itself most
prominently near the kinematic limit $(x_1, x_2, y_a, y_b) \rightarrow 1$ 
of the inclusive reaction.
The entropy of the constituent configurations in this region bears information on fractal characteristics of hadron interactions at small scales. 
In the vicinity of the kinematic limit (i.e. near the
fractal limit $\Omega^{-1} \rightarrow \infty$), the momentum dependent part 
\begin{equation}
S_{\Omega} = \delta_1\ln{(1\!-\!x_1)} + \delta_2\ln{(1\!-\!x_2)} +
\varepsilon_a\ln{(1\!-\!y_a)} + \varepsilon_b\ln{(1\!-\!y_b)} + \ln{\Omega_0}
\label{eq:17}
\end{equation}
of the entropy (\ref{eq:10}) 
can be expressed in the form that admit physical interpretation 
of quantization of the fractal dimensions $\delta_1$, $\delta_2$, $\varepsilon_a$
and $\varepsilon_b$.

The maximization of $S_{\Omega}$ constrained by condition (\ref{eq:3}) 
gives the asymptotic formulae \cite{3}
\begin{equation}
1\!-\!x_1 = h_1(p)
\frac{\delta_1}{\delta\!+\varepsilon},
\ \ \ \ \ \ \ \
1\!-\!x_2 = h_2(p)
\frac{\delta_2}{\delta\!+\varepsilon},
\label{eq:18}
\end{equation}
\begin{equation}
1\!-\!y_a = h_a(p)
\frac{\varepsilon_a}{\delta\!+\varepsilon},
\ \ \ \ \ \ \ \
1\!-\!y_b = h_b(p)
\frac{\varepsilon_b}{\delta\!+\varepsilon},
\label{eq:19}
\end{equation} 
valid near the kinematic boundary.
The functions $h_1(p)$, $h_2(p)$, $h_a(p)$, $h_b(p)$   
depend explicitly on the momentum $p$ of the inclusive particle, 
$\delta \equiv \delta_1+\delta_2$ and $\epsilon  \equiv \epsilon_a+\epsilon_b$.
Using expressions (\ref{eq:18}) and (\ref{eq:19}), the entropy  (\ref{eq:17})
can be decomposed as follows
\begin{equation}
S_{\Omega} = S_{\Upsilon}-S_{\Gamma}+ \ln\Omega_{0}.
\label{eq:20}
\end{equation} 
The first term, $S_{\Upsilon}$, describes the dependence of the entropy 
on the momenta and masses of the colliding and inclusive particles, 
the second term, 
\begin{equation}
S_{\Gamma} = (\delta+\epsilon) \ln (\delta+\epsilon)
-\delta_1  \ln \delta_1
-\delta_2  \ln \delta_2
-\epsilon_a  \ln \epsilon_a
-\epsilon_b \ln \epsilon_b,
\label{eq:21}
\end{equation} 
depends solely on fractal dimensions,
and the third one is a constant that guaranties normalization.
The formula (\ref{eq:21}) allows us to derive physical consequences provided 
that the fractal dimensions are expressed as integer multiples of the same constant $d$,
\begin{equation}
\delta_{1} = n_{\delta_1}\cdot d, \ \ \
\delta_{2} = n_{\delta_2}\cdot d, \ \ \
\epsilon_{a} =  n_{\epsilon_a}\cdot d, \ \ \
\epsilon_{b} =  n_{\epsilon_b}\cdot d .
\label{eq:22}
\end{equation}

In that case, the entropy $S_{\Gamma}$ can be interpreted within a statistical 
ensemble of fractal
configurations of the internal structures of the colliding hadrons (or nuclei) and fractal
configurations corresponding to the fragmentation processes in the final state.
The statistical ensemble 
is considered as a collection of $n_{\delta_1}$ fractals with 
random configurations but with the same fractal dimension $\delta_1$, together with
an analogous set of $n_{\delta_2}$ interacting fractals
with the fractal dimension $\delta_2$, which are
combined via binary sub-processes with the 
collection of $n_{\varepsilon_a}$ fractals with 
random configurations but with the same fractal dimension $\varepsilon_a$  in the final state, 
and the corresponding set of $n_{\varepsilon_b}$ fractals 
with the fractal dimension $\varepsilon_b$. 
For large numbers of the configurations, $S_{\Gamma}$ 
can be rewritten as follows
\begin{equation}
S_{\Gamma} = d \cdot \ln({\Gamma}_{\delta_1,\delta_2,\epsilon_a,\epsilon_b}),
\label{eq:23}
\end{equation}
where
\begin{equation}
{\Gamma}_{\delta_1,\delta_2,\epsilon_a,\epsilon_b} \equiv
\frac{(n_{\delta_1}+n_{\delta_2}+ n_{\epsilon_a}+n_{\epsilon_b})!}
{ n_{\delta_1}!\cdot n_{\delta_2}! \cdot n_{\epsilon_a}!\cdot  n_{\epsilon_b}!    }
=
{\Gamma}_{\delta,\epsilon}\cdot
{\Gamma}_{\delta_1,\delta_2}\cdot
{\Gamma}_{\epsilon_a,\epsilon_b} 
\label{eq:24}
\end{equation}
and 
\begin{equation}
{\Gamma}_{\delta,\epsilon}=\frac{(n_{\delta}+n_{\epsilon})!} {n_{\epsilon}!\cdot n_{\delta}!     }, \ \  \
{\Gamma}_{\delta_1,\delta_2}=\frac{(n_{\delta_1}+n_{\delta_2})!} {n_{\delta_1}!\cdot n_{\delta_2}!},   \ \ \   
{\Gamma}_{\epsilon_a,\epsilon_b}=\frac{(n_{\epsilon_a}+n_{\epsilon_b})!} {n_{\epsilon_a}!\cdot n_{\epsilon_b}!     }.
\label{eq:25}
\end{equation}

According to statistical physics, the formulae (\ref{eq:23}) - (\ref{eq:25}) give us the possibility to
interpret the entropy $S_{\Gamma}$, expressed in units of the dimensional quantum $d$, as the logarithm of the number of different ways,
${\Gamma}_{\delta_1,\delta_2,\epsilon_a,\epsilon_b}$,
in which the fractal
dimensions of the interacting fractal structures can be composed from the identical
dimensional quanta, each of the size~$d$. 
The symbol  ${\Gamma}_{\delta,\epsilon}$ represents the  number of ways 
how the overall number $n=n_{\delta}+n_{\epsilon}$ of the dimensional quanta can be shared among $n_{\delta}$ portions pertaining to the fractal
dimensions of the colliding objects and $n_{\epsilon}$ portions belonging 
to the fractal dimensions characterizing fractal structures of the fragmentation processes.
The symbols  ${\Gamma}_{\delta_1,\delta_2}$ and ${\Gamma}_{\epsilon_a,\epsilon_b}$ 
represent the numbers of 
ways in which the corresponding  numbers $n_{\delta}=n_{\delta_1} + n_{\delta_2}$ 
and $n_{\epsilon}=n_{\epsilon_a} + n_{\epsilon_a}$ 
of the dimensional quanta can be distributed between the fractal dimensions of
the single fractals in the initial and final states, respectively.

The statistical interpretation of the entropy (\ref{eq:21}) is only possible under the quantization of fractal dimensions (\ref{eq:22}),
where $d $ is an elementary dimensional quantum. The quantization is a kind of ordering
that diminishes the total entropy ($S_{\Gamma}$ enters with minus sign into Eq.(\ref{eq:20})).

\subsection{Conservation of the number of cumulativity quanta}

The quantization of the fractal dimension, $D =  n_D \cdot d$, is connected
with quantum character of the fractal cumulativity $C(D,\zeta)$. 
Using expression (\ref{eq:13}), we can write
this quantity in the form
\begin{equation}
C = n_C \cdot d , \ \ \ \ \ \
n_C(n_D,\zeta)=n_D\cdot \frac{\zeta}{1-\zeta},
\label{eq:26}
\end{equation}
where $n_C(n_D,\zeta)$ represents the number of quanta of the fractal cumulativity 
expressed in units of the dimensional quantum $d$.
The quantum character of the fractal dimensions has
profound impact on the physical content of the conservation law for the fractal cumulativity.
According to (\ref{eq:12}), 
the number of the cumulativity quanta is conserved at any resolution
given by arbitrary momenta $P_1, P_2$ and $p$ of the colliding and inclusive particles. 
The conservation law can be formulated  as follows:

{\it The number of quanta of fractal cumulativity 
before a constituent interaction  is equal to 
the number of quanta of fractal cumulativity 
after the constituent interaction  for any binary sub-process}, 

\begin{equation}
 \sum_i^{in} n_C(n_{D_i},\zeta_i) = \sum_j^{out} n_C(n_{D_j},\zeta_j).
\label{eq:27}
\end{equation}
The quantization of the dimension $D$ and cumulativity $C(D,\zeta)$ 
is based on the assumptions of the fractal 
self-similarity of internal hadron structure, fractal nature of fragmentation process,
and locality of hadron interactions at a constituent level up to the kinematic limit.

The conservation law for the quanta of the fractal cumulativity follows 
from general physical principles.  
According to the Noether's
theorem, for every conservation law of a continuous quantity 
there must be a continuous symmetry. 
In our case, the corresponding
symmetry is a scale-dependent translation symmetry which guaranties
the conservation law for the fractal cumulativity at any scale fixed by the minimal
necessary level of resolution.

\section{Conclusions}
In summary we conclude that
$z$-scaling is a specific feature of high-$p_T$ particle production established
in $p+p$ and $\bar p+p$ collisions at the U70, ISR, S$\rm {p\bar p}$S, Tevatron, RHIC and LHC.
It reflects the self-similarity, locality, and fractality of hadron interactions
at a constituent level.
The scaling behavior was confirmed for inclusive production of different hadrons, jets, 
heavy quarkonia and top quark.
The hypothesis of the self-similarity and fractality was tested in $Au+Au$ collisions
at RHIC using $z$-presentation of spectra of negative hadrons.
The analysis of the STAR BES-I data indicates  
energy and multiplicity independence of the scaling function $\psi(z)$.
The variable $z$ depends on the multiplicity density, "heat capacity", and 
entropy of constituent configurations of the interacting system.
The constituent energy loss as a function of the energy and centrality of collisions 
and the transverse momentum of inclusive particles was estimated.

We have shown that $z$-scaling, containing the principle 
of maximum entropy, includes a conservation law of the "fractal cumulativity"  $C(D,\zeta)$. 
This quantity reflects the ability of the fractal systems to create structural constituents
with certain  degree of local compactness. 
The cumulativity of a fractal object or a fractal-like process with fractal dimension $D$
carried by its constituent with the momentum fraction $\zeta$ is 
proportional to the dimension $D$ and represents a simple function of $\zeta$.
It was shown that a composition rule for $C(D,\zeta)$ connects 
the fractal cumulativity at different scales.
The fractal dimension $D$ is interpreted as a quantity which has quantum nature.
The quantization of fractal dimensions results in preservation of the number 
of the cumulativity quanta $n_C(n_D,\zeta)$ in binary sub-processes at any resolution.

\section*{Acknowledgements}
The investigation of (I.Z.) was supported by the
RVO61389005 institutional support 
and by the MEYS of the Czech Republic under the contracts LTT18021 and LTT17018.

\end{document}